\documentclass[preprint]{iucr}  % DO NOT 
\RequirePackage{graphicx} %DELETE THIS LINE

     \journalcode{J}              % Indicate the journal to which submitted
\usepackage{amssymb}
\usepackage{amsmath}
\usepackage{xcolor}

\begin{document}  % DO NOT DELETE THIS LINE

     %-------------------------------------------------------------------------
     % The introductory (header) part of the paper
     %-------------------------------------------------------------------------

     % The title of the paper. Use \shorttitle to indicate an abbreviated title
     % for use in running heads (you will need to uncomment it).

\title{Desmearing two-dimensional small-angle neutron scattering data by central moment expansions}

     % Authors' names and addresses. Use \cauthor for the main (contact) author.
     % Use \author for all other authors. Use \aff for authors' affiliations.
     % Use lower-case letters in square brackets to link authors to their
     % affiliations; if there is only one affiliation address, remove the [a].

\cauthor[a,b]{Guan-Rong}{Huang}{huangrn@ess.nthu.edu.tw}

\author[c]{Chi-Huan}{Tung}
\author[d]{Weijian}{Hua}
\author[d]{Yifei}{Jin}
\author[e]{Lionel}{Porcar}
\author[f]{Yuya}{Shinohara}
\author[g]{Christoph U.}{Wildgruber}
\author[c]{Changwoo}{Do}
\cauthor[c]{Wei-Ren}{Chen}{chenw@ornl.gov}

\aff[a]{Department of Engineering and System Science, National Tsing Hua University, Hsinchu 30013, \country{Taiwan}}
\aff[b]{Physics Division, National Center for Theoretical Sciences, Taipei 10617, \country{Taiwan}}
\aff[c]{Neutron Scattering Division, Oak Ridge National Laboratory, Oak Ridge, TN 37831, \country{United States}}
\aff[d]{Department of Mechanical Engineering, University of Nevada, Reno, NV 89557, \country{United States}}\aff[d]{Institut Laue-Langevin, B.P. 156, F-38042 Grenoble Cedex 9, \country{France}}
\aff[e]{Materials Science and Technology Division, Oak Ridge National Laboratory, Oak Ridge, TN 37831, \country{United States}}
\aff[f]{Neutron Technologies Division, Oak Ridge National Laboratory, Oak Ridge, TN 37831, \country{United States}}

\maketitle   % DO NOT DELETE THIS LINE

\begin{synopsis}

This paper introduces a novel desmearing methodology utilizing the central moment expansion technique to correct resolution smearing in two-dimensional SANS data, particularly for anisotropic patterns in rheo-SANS experiments. This model-free approach accurately reconstructs true intensity distributions, as validated through computational benchmarks and experimental data from shear-induced crystalline structures in an aqueous solution of nanocomposites. The methodology effectively recovers structural features and quantifies experimental uncertainties, enhancing the quantitative analysis of SANS data for complex fluid systems.

\end{synopsis}

\clearpage

\begin{abstract}

Resolution smearing is a critical challenge in the quantitative analysis of two-dimensional small-angle neutron scattering (SANS) data, particularly in studies of soft matter flow and deformation using SANS. We present the central moment expansion technique to address smearing in anisotropic scattering spectra, offering a model-free desmearing methodology. By accounting for directional variations in resolution smearing and enhancing computational efficiency, this approach reconstructs desmeared intensity distributions from smeared experimental data. Computational benchmarks using interacting hard-sphere fluids and Gaussian chain models validate the accuracy of the method, while simulated noise analyses confirm its robustness under experimental conditions. Furthermore, experimental validation using the rheo-SANS data of shear-induced micellar structures demonstrates the practicality and effectiveness of the proposed algorithm. The desmearing technique provides a powerful tool for advancing the quantitative analysis of anisotropic scattering patterns, enabling precise insights into the interplay between material microstructure and macroscopic flow behavior.

\end{abstract}

\clearpage

     %-------------------------------------------------------------------------
     % The main body of the paper
     %-------------------------------------------------------------------------

\section{Introduction}
\label{sec:Introduction}

Since the 1980s \cite{Hayter1984, Hayter1987}, rheo-small-angle neutron scattering (rheo-SANS) has emerged as a powerful technique that integrates rheological measurements with SANS to investigate the interplay between the macroscopic flow properties of materials and their underlying microstructural distortions. By subjecting materials to mechanical shear or extensional flow while simultaneously probing their mesoscale structures, rheo-SANS offers unique insights into the dynamic behavior of complex fluids, such as polymer solutions, colloidal suspensions, and wormlike micelles. The anisotropic scattering patterns observed under flow provide critical information about structural alignment, deformation, and relaxation mechanisms.

Nevertheless, the measured scattering spectra are inherently smeared by instrumental resolution, which is determined by factors such as wavelength spread, beam size, beam divergence, and detector pixel size. This resolution smearing distorts the measured scattering features, complicating the quantitative extraction of structural parameters.

To establish a robust connection between observed rheological properties and microstructural transformations, it is essential to address the effects of instrumental resolution. While significant progress has been made in interpreting anisotropic scattering data and desmearing of isotropic scattering data, the direct desmearing of resolution effects on anisotropic scattering spectra remains a critical yet under-explored challenge in quantitative analysis.

Recently, we introduced a robust methodology for desmearing azimuthally-averaged one-dimensional (1D) small-angle scattering intensity profiles using the central moment expansion technique \cite{Huang2023}. This approach eliminates the need for predefined model assumptions, enabling model-free desmearing and providing a reliable framework to correct instrumental resolution effects. The methodology demonstrated excellent numerical accuracy and was validated against computational benchmarks, offering a practical solution for analyzing isotropic scattering data.

Building upon this framework, the present work extends the central moment expansion technique to address the desmearing of two-dimensional (2D) anisotropic scattering spectra. This extension is particularly significant for rheo-SANS experiments, where anisotropic scattering patterns frequently arise due to structural distortions induced by flow. Desmearing 2D spectra introduces unique challenges, including the need to account for directional variations in resolution smearing and the increased computational complexity associated with higher-dimensional data.

Our goal is to develop a generalized methodology capable of desmearing intensity distributions of 2D spectra without relying on predefined structural models. This approach aims to enhance the quantitative analysis of anisotropic scattering data, enabling a more precise determination of structural parameters. Ultimately, this work seeks to deepen our understanding of the intricate relationship between rheology and microstructural distortions in flowing materials, advancing the potential of rheo-SANS in exploring complex fluid systems.

\section{Method}
\label{sec:Method}

The measured scattering intensity \(I_{\text{exp}}(Q_x,Q_y)\) is always smeared by the instrumental resolution function \( R(Q_x, Q_y) \), where \(Q_x\) and \(Q_y\) represent the momentum transfer between incident and scattered neutrons along horizontal (\( x \)-axis) and vertical (\( y \)-axis) directions, respectively. The \( y \)-axis is set to be oriented anti-parallel to the direction of gravitational acceleration. The resolutions on the detector plane along vertical and horizontal directions vary at different points of \( Q_x\) and \( Q_y\). Assuming the scattering intensity before the smearing is \( I(Q_x,Q_y)\), \(I_{\text{exp}}(Q_x,Q_y)\) can be modeled by the following two-dimensional convolution integral \cite{Born_Wolf}:
\begin{equation}
    \label{eq:1}
    I_{\text{exp}}(Q_x,Q_y) = \int_{-\infty}^{\infty} \int_{-\infty}^{\infty}\mathrm{d}x \,\mathrm{d}y \, I(x,y) R(Q_x-x,Q_y-y).
\end{equation}
The smearing effect of \( R(Q_x, Q_y) \) is generally modeled as a 2D Gaussian function:
%---------------------------
\begin{equation}
    \label{eq:2}
    R(Q_x, Q_y) = \frac{1}{2\pi} \exp{\left(-\frac{Q_x^2}{2\sigma_{Q_x}^2}-\frac{Q_y^2}{2\sigma_{Q_y}^2}\right)},
\end{equation}
%---------------------------
where \( \sigma_{Q_x} \) and \( \sigma_{Q_y} \) are the standard deviations parallel to the \( Q_x\) and \( Q_y\) directions, respectively. When \( \sigma_{Q_x} \) and \( \sigma_{Q_y} \) are smaller enough than the spacing between pixels, the major contributions in Eqn.~\eqref{eq:1} come from the central moments of \( R(Q_x, Q_y) \), defined as
\begin{align}
    \label{eq:3}
    \langle Q_x^n \rangle &= \int_{-\infty}^{\infty} \int_{-\infty}^{\infty}\mathrm{d}x \,\mathrm{d}y \, (x-Q_x)^n R(Q_x-x,Q_y-y), \nonumber \\
    \langle Q_y^n \rangle &= \int_{-\infty}^{\infty} \int_{-\infty}^{\infty}\mathrm{d}x \,\mathrm{d}y \, (y-Q_y)^n R(Q_x-x,Q_y-y).
\end{align}
Because the functional form of the resolution function in Eqn.~\eqref{eq:2} is symmetrical to the mean values: \( Q_x\) and \(Q_y\), the corresponding central moments for the cases of odd \( n \) values must vanish. The results of Eqn.~\eqref{eq:3} are simply related to the powers of \( \sigma_{Q_x} \) and \( \sigma_{Q_y} \) as:
\begin{equation}
    \label{eq:4}
    \langle Q_x^n \rangle = 
    \begin{cases}
        0, & \text{if \(n \) is odd} \\
        \sigma_{Q_x}^n (n-1)!!, & \text{otherwise}
    \end{cases},
    \quad
    \langle Q_y^n \rangle = 
    \begin{cases}
        0, & \text{if \(n \) is odd} \\
        \sigma_{Q_y}^n (n-1)!!, & \text{otherwise}
    \end{cases},
\end{equation}
where \( !! \) denotes the double factorial. To investigate how these moments affect the smearing of ground truth intensity, we can expand \( I(x,y) \) in Eqn.~\eqref{eq:1} using the Taylor expansion of two variables at \( (Q_x,Q_y)\) as:
\begin{equation}
    \label{eq:5}
    I(x,y) = \sum_{l=0}^{\infty}\sum_{m=0}^{\infty} \frac{(x-Q_x)^l(y-Q_y)^m}{l!m!}\frac{\partial^{l+m}}{\partial Q_x^l \partial Q_y^m }I(Q_x,Q_y).
\end{equation}
As a result, by using Eqs.~\eqref{eq:4} and~\eqref{eq:5}, Eqn.~\eqref{eq:1} can be simplified as
\begin{equation}
\label{eq:6}
    I_{\text{exp}}(Q_x,Q_y) = \sum_{l,m=0,2,4,\ldots}^{\infty} \frac{\sigma_{Q_x}^l\sigma_{Q_y}^m}{2^{(l+m)/2}(l/2)!(m/2)!}\frac{\partial^{l+m}}{\partial Q_x^l \partial Q_y^m }I(Q_x,Q_y).
\end{equation}
When \(\frac{\sigma_{Q_x}^l\sigma_{Q_y}^m}{2^{(l+m)/2}(l/2)!(m/2)!} \) is smaller than the magnitude of \( \frac{\partial^{l+m}}{\partial Q_x^l \partial Q_y^m }I(Q_x,Q_y) \), Eqn.~\eqref{eq:6} can be approximated as
\begin{equation}
\label{eq:7}
    I_{\text{exp}}(Q_x,Q_y) \simeq \left(1 + \frac{\sigma_{Q_x}^2}{2}\frac{\partial^2}{\partial Q_x^2} + \frac{\sigma_{Q_y}^2}{2}\frac{\partial^2}{\partial Q_y^2}\right) I(Q_x,Q_y).
\end{equation}
As a result, the relationship between \( I(Q_x,Q_y) \) and \( I_{\text{exp}}(Q_x,Q_y) \) can be approximated by:
\begin{align}
\label{eq:8}
I(Q_x,Q_y) &\simeq \left(1 + \frac{\sigma_{Q_x}^2}{2}\frac{\partial^2}{\partial Q_x^2} + \frac{\sigma_{Q_y}^2}{2}\frac{\partial^2}{\partial Q_y^2}\right)^{-1} I_{\text{exp}}(Q_x,Q_y) \nonumber \\
&= \left(1 - \frac{\sigma_{Q_x}^2}{2}\frac{\partial^2}{\partial Q_x^2} - \frac{\sigma_{Q_y}^2}{2}\frac{\partial^2}{\partial Q_y^2}\right) I_{\text{exp}}(Q_x,Q_y) + O(\sigma_{Q_x}^4,\sigma_{Q_y}^4,\sigma_{Q_x}^2\sigma_{Q_y}^2).
\end{align}
When \( \sigma_{Q_x} \to 0 \) and \( \sigma_{Q_y} \to 0 \), \( I(Q_x, Q_y) = I_{\text{exp}}(Q_x, Q_y) \), indicating the absence of smearing effects due to instrumental resolution. In the following section, we will explore how the magnitudes of \( \sigma_{Q_x} \) and \( \sigma_{Q_y} \) and their relative ratios influence the experimental scattering data. We will calculate \( I(Q_x, Q_y) \) using Eqn.~\eqref{eq:8} and compare it to numerical simulations of Gaussian chain polymers in both quiescent and deformed states. This method will also be applied to SANS data of shear-induced micellar order structures.

%---------------------------------------------
\begin{figure}
\centerline{
  \includegraphics[width =\columnwidth]{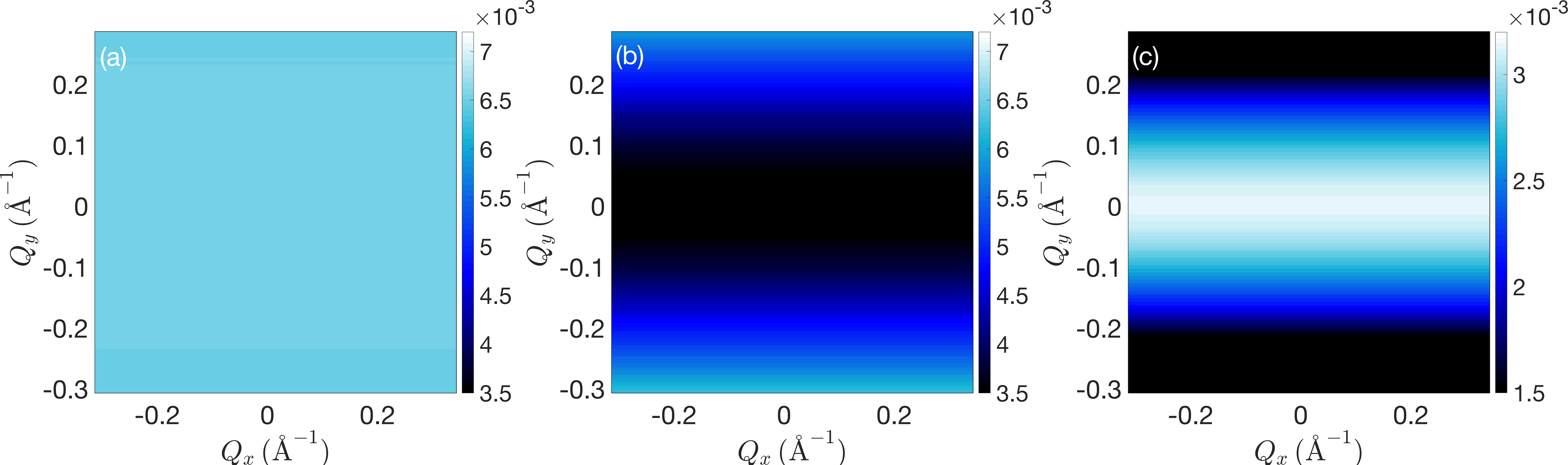}
  %\vspace{0 cm}
}  
\caption{(a) The \( \sigma_{Q_x} \) map on the \( Q_x\)-\( Q_y \) plane, showing a magnitude of approximately \( 7.0 \times 10^{-3} \, \rm{\AA^{-1}} \) with an essentially isotropic distribution. (b) The \( \sigma_{Q_y} \) map on the same plane, where the magnitude ranges from \( 3.5 \times 10^{-3} \, \rm{\AA^{-1}} \) to \( 7.0 \times 10^{-3} \, \rm{\AA^{-1}} \), exhibiting a clear anisotropy along the \( Q_y \) direction. (c) The convex nature of the resolution function along the \( Q_y \) axis is highlighted by the difference between (a) and (b).}
\label{fig:1}
\end{figure}
%---------------------------------------------

Before evaluating the efficacy of the proposed desmearing algorithm, we examine the general characteristics of the SANS resolution function described in Eqn.~\eqref{eq:2}. As an illustrative example, we consider the instrument resolution of the Extended Q-Range Small-Angle Neutron Scattering Diffractometer (EQSANS) at the Spallation Neutron Source (SNS). Figure~\ref{fig:1}(a) displays the \( \sigma_{Q_x} \) map on the \( Q_x\)-\( Q_y \) plane, showing a magnitude of approximately \( 7.0 \times 10^{-3} \, \rm{\AA^{-1}} \) with an essentially isotropic distribution. In contrast, Figure~\ref{fig:1}(b) presents the \( \sigma_{Q_y} \) map on the same plane, where the magnitude ranges from \( 3.5 \times 10^{-3} \, \rm{\AA^{-1}} \) to \( 7.0 \times 10^{-3} \, \rm{\AA^{-1}} \). A clear anisotropy is observed along the \( Q_y \) direction. The convex nature of the resolution function along the \( Q_y \) axis is further highlighted by the difference between the results shown in Figures~\ref{fig:1}(a) and \ref{fig:1}(b), as depicted in Figure~\ref{fig:1}(c).

\section{Computational Benchmark}
\label{sec:Computational Benchmark}

In this section, we assess the numerical accuracy of the desmearing technique, building upon the mathematical framework introduced above. To evaluate the performance, we use interacting monodisperse hard-sphere fluids as a benchmark system. The coherent neutron scattering intensity, \( I(Q) \), for this system is expressed in terms of the intra-particle form factor \( P(Q) \) and the inter-particle structure factor \( S(Q) \) as follows:
%-------------------------------
\begin{equation}
I(Q) = n \Delta \rho^2 v^2 P(Q) S(Q),
\label{eq:3.1}
\end{equation}
%-----------------------------------------
where \( n \) represents the particle number density, and \( \Delta \rho \) denotes the excess scattering length density. The volume of a hard sphere, \( v \), is given by \( v = \frac{4\pi}{3} R^3 \), where \( R \) is the radius of the sphere.

The intra-particle form factor, \( P(Q) \), is defined as:
%-------------------------------
\begin{equation}
P(Q) = \left[3\frac{\sin(QR) - QR\cos(QR)}{(QR)^3}\right]^2.
\label{eq:3.2}
\end{equation}
%--------------------------------

The inter-particle structure factor, \( S(Q) \), is determined by solving the Ornstein-Zernike equation~\cite{OZ} with the Percus-Yevick closure~\cite{PY}. This computation is performed using Baxter's analytical method~\cite{Baxter1, Baxter2, Hansen}, enhanced by the Wertheim correction~\cite{Wertheimt} to account for the structural collectivity.

\begin{figure}
\centerline{
  \includegraphics[width =\columnwidth]{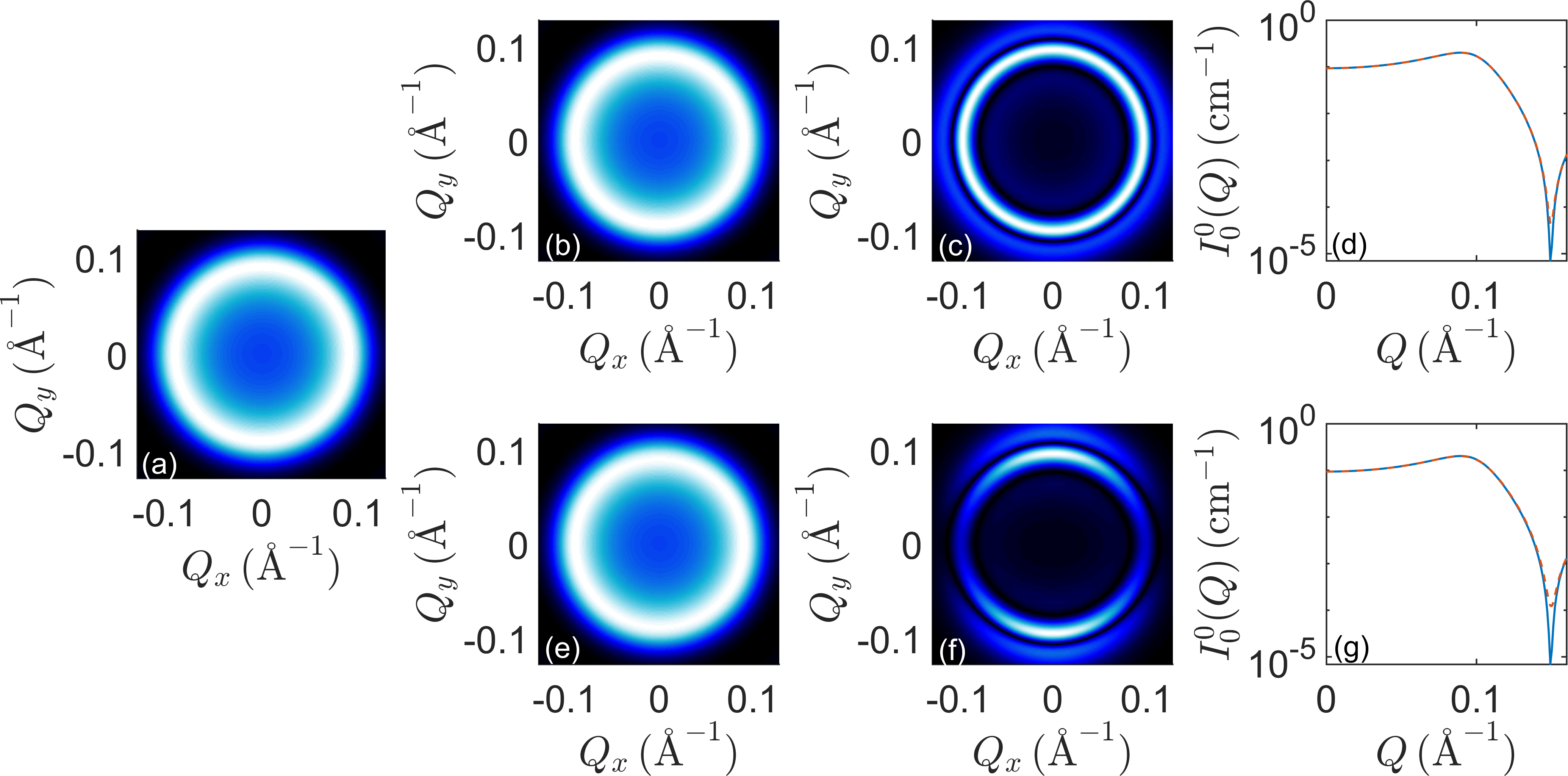}
  %\vspace{0 cm}
}  
\caption{
(a) Theoretical two-dimensional intensity, \( I(\textbf{Q}) \), for a hard-sphere fluid (\( \phi = 0.3 \)).  
(b) Intensity smeared by the isotropic resolution function, \( I^I_S(\textbf{Q}) \).  
(c) Difference, \( \Delta I \), between \( I^I_S(\textbf{Q}) \) and \( I(\textbf{Q}) \), showing angularly isotropic distortions.  
(d) Angularly averaged intensity, \( I^I_S(Q) \), after isotropic smearing.  
(e) Intensity smeared by the anisotropic resolution function, \( I^A_S(\textbf{Q}) \).  
(f) Difference between \( I^A_S(\textbf{Q}) \) and \( I(\textbf{Q}) \), revealing angular anisotropy in distortions.  
(g) Comparison of the isotropic component \( I_0^0(Q) \), extracted from \( I^A_S(\textbf{Q}) \), with \( I^I_S(Q) \), highlighting more pronounced distortions caused by anisotropic smearing. For the isotropic resolution, \( dQ_x = dQ_y = 1.0 \times 10^{-3} \,\text{\AA}^{-1} \). For the anisotropic resolution, \( dQ_x = 1.5 \times 10^{-3} \,\text{\AA}^{-1} \) and \( dQ_y = 6.0 \times 10^{-3} \,\text{\AA}^{-1} \).
}
\label{fig:2}
\end{figure}

The resulting scattering intensities are presented in Fig.~\ref{fig:2}. Panel (a) shows the theoretically calculated two-dimensional intensity, \( I(\textbf{Q}) \), for a hard-sphere fluid with a volume fraction \( \phi = 0.3 \). Panel (b) displays the intensity smeared by the isotropic resolution function, \( I^I_S(\textbf{Q}) \), which is obtained by convolving \( I(\textbf{Q}) \) with the isotropic resolution function illustrated in Fig.~\ref{fig:1}(a). The superscript \( I \) and subscript \( S \) indicate that the spectrum has been smeared using the isotropic resolution function.

Although \( I(\textbf{Q}) \) and \( I^I_S(\textbf{Q}) \) appear nearly identical upon visual inspection, a pixel-wise subtraction reveals spectral distortions induced by the smearing process. This difference, which highlights the intensity distortion, denoted as \( \Delta I \), is shown in Fig.~\ref{fig:2}(c). Notably, the distortion remains angularly isotropic.

Next, \( I(\textbf{Q}) \) is smeared by the anisotropic resolution function, depicted in Fig.~\ref{fig:1}, yielding the intensity \( I^A_S(\textbf{Q}) \), as shown in Fig.~\ref{fig:2}(e). Although no visual difference is discernible between \( I(\textbf{Q}) \) and \( I^A_S(\textbf{Q}) \), a pixel-wise subtraction reveals the spectral distortions introduced by the anisotropic smearing process. This difference, highlighting the intensity distortion, is shown in Fig.~\ref{fig:2}(f). In contrast to the angularly isotropic \( \Delta I \) in Fig.~\ref{fig:2}(c), the anisotropic resolution function introduces a clear angular anisotropy in \( I^A_S(\textbf{Q}) \). While this anisotropy is not visually apparent in Fig.~\ref{fig:2}(e), it becomes evident upon subtraction.

%---------------------------------------------
%---------------------------------------------

The anisotropic scattering intensity, \( I^A_S(\textbf{Q}) \), can be expanded as a linear combination of real spherical harmonics (RSH):
\begin{equation}
    I^A_S(\textbf{Q}) = \sum_{l,m} I_l^m(Q) Y_l^m(\theta, \phi),
\label{eq:3.3}
\end{equation}
where \( Y_l^m(\theta, \phi) \) denotes the RSH of degree \( l \) and order \( m \), with \( \theta \) and \( \phi \) representing the polar and azimuthal angles in reciprocal space, respectively. Among the RSH components of \( I^A_S(\textbf{Q}) \), the isotropic component, \( I_0^0(Q) \), is most relevant for analyzing resolution-induced spectral distortion, as it provides the angularly averaged scattering intensity. 

The comparison of \( I_0^0(Q) \), extracted from \( I^A_S(\textbf{Q}) \) using Eq.~\eqref{eq:3.3}, with \( I(Q) \) is shown in Fig.~\ref{fig:2}(g). By comparing the first local minimum of \( I_0^0(Q) \) with that of \( I^I_S(Q) \) in Fig.~\ref{fig:2}(d), it becomes evident that the anisotropic resolution function not only introduces spectral anisotropy but also causes more pronounced distortions in the angularly averaged intensity.

To evaluate the effectiveness of the proposed two-dimensional desmearing algorithm, the smeared intensities, \( I^I_S(\textbf{Q}) \) and \( I^A_S(\textbf{Q}) \), shown in Figs.~\ref{fig:2}(b) and \ref{fig:2}(e), respectively, are processed using the central moment expansion method described in Section~\ref{sec:Method}, where the spacing between adjacent \( Q_x \) or \( Q_y \) points is \( 1 \times 10^{-3} \,\text{\AA}^{-1} \). The resulting desmeared intensities are presented in Figs.~\ref{fig:2}(b) and \ref{fig:2}(e). 

%---------------------------------------------
\begin{figure}
\centerline{
  \includegraphics[width =\columnwidth]{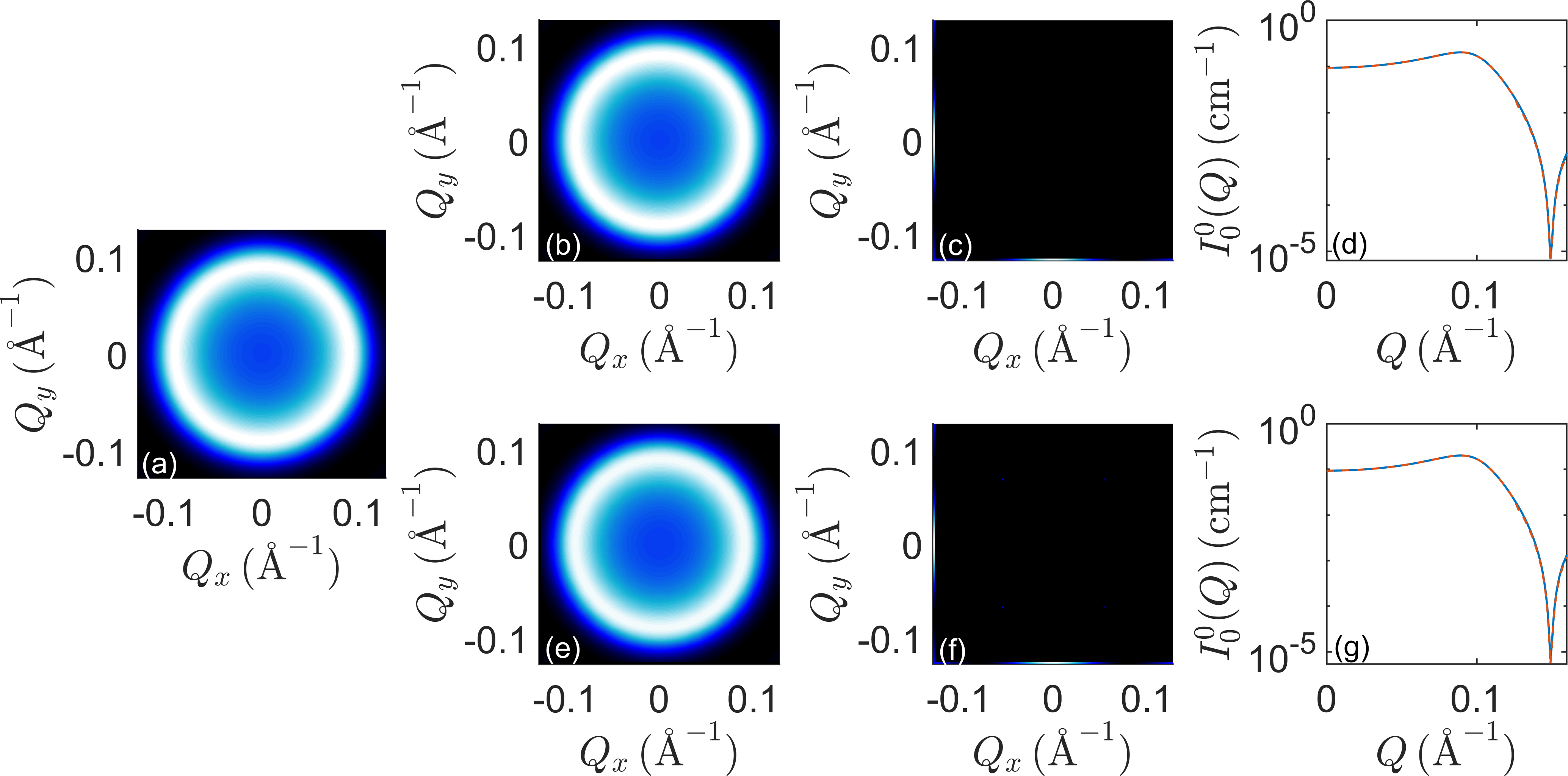}
  %\vspace{0 cm}
}  
\caption{Panels (b) and (d) show the desmeared intensities obtained using the central moment expansion method applied to the smeared data (Figs.~\ref{fig:2}(b) and \ref{fig:2}(e)). The intensity differences relative to the ground truth unsmeared intensity, \( I(\textbf{Q}) \) (Panel (a)), are displayed in Panels (c) and (f). Panels (d) and (g) compare the angularly averaged desmeared intensity, \( I_0^0(Q) \), highlighting the algorithm's effectiveness in resolving spectral anisotropy and accurately recovering the angularly averaged intensity. For the isotropic resolution, \( dQ_x = dQ_y = 1.0 \times 10^{-3} \,\text{\AA}^{-1} \). For the anisotropic resolution, \( dQ_x = 1.5 \times 10^{-3} \,\text{\AA}^{-1} \) and \( dQ_y = 6.0 \times 10^{-3} \,\text{\AA}^{-1} \).}
\label{fig:3}
\end{figure}
%---------------------------------------------

The differences between the desmeared intensities and the ground truth unsmeared intensity, \( I(\textbf{Q}) \), shown in Figs.~\ref{fig:2}(a), are illustrated in Figs.~\ref{fig:2}(c) and \ref{fig:2}(f). Additionally, a comparison of the angularly averaged desmeared intensity, \( I_0^0(Q) \), is provided in Figs.~\ref{fig:2}(d) and \ref{fig:2}(g). These results clearly demonstrate that the proposed desmearing algorithm effectively resolves artificial features of spectral anisotropy while achieving high numerical accuracy in retrieving the angularly averaged intensity.
   
\begin{figure}
\centerline{
  \includegraphics[width =\columnwidth]{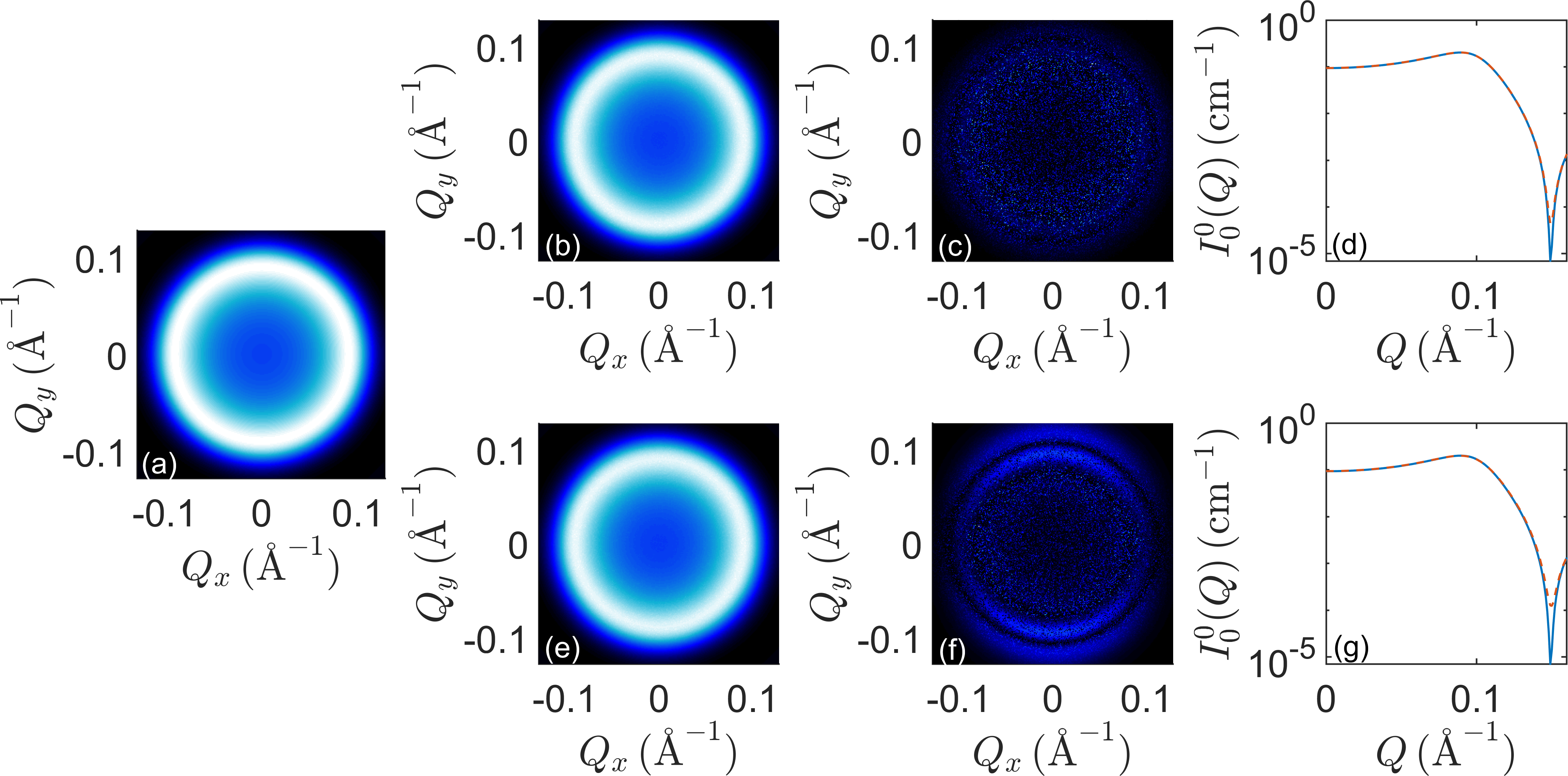}
  %\vspace{0 cm}
}  
\caption{Gaussian noise, representing statistical errors at 1\% of the absolute intensity, is introduced to each pixel to simulate experimental data. Panel~(a) shows the two-dimensional scattering intensity with noise, while Panels~(b) and~(e) present intensities smeared by isotropic and anisotropic resolution functions, respectively. Comparison with the ground truth $I(\mathbf{Q})$ (Figs.~\ref{fig:4}(c) and~(f)) reveals that intensity fluctuations, alongside resolution effects, contribute to the smearing. Fig.~\ref{fig:4}(g) shows that the anisotropic resolution induces more smearing than the isotropic resolution (Fig.~\ref{fig:3}(g)). For the isotropic resolution, \( dQ_x = dQ_y = 1.0 \times 10^{-3} \,\text{\AA}^{-1} \). For the anisotropic resolution, \( dQ_x = 1.5 \times 10^{-3} \,\text{\AA}^{-1} \) and \( dQ_y = 6.0 \times 10^{-3} \,\text{\AA}^{-1} \).}
\label{fig:4}
\end{figure}

To more accurately simulate experimental data, the uncertainties inherent in scattering intensity measurements are taken into account. Gaussian noise, with a magnitude of approximately 1\% of the absolute intensity, is introduced to each pixel to represent statistical errors, in line with typical SANS studies. The results are presented in Fig.~\ref{fig:4}. Panel~(a) displays the two-dimensional scattering intensity, with each pixel incorporating Gaussian noise. Panels~(b) and~(e) show the scattering intensities smeared by isotropic and anisotropic instrument resolution functions, respectively.

%---------------------------------------------
%---------------------------------------------

In comparison to the results presented in Figs.~\ref{fig:3}(c) and~(g), the differences between the smeared intensities and the ground truth $I(\mathbf{Q})$, as shown in Figs.~\ref{fig:4}(c) and~(f), clearly illustrate that, in addition to the effects of instrument resolution, intensity fluctuations contribute to the smearing of the measured scattering intensity. As observed in Fig.~\ref{fig:3}(g), Fig.~\ref{fig:4}(g) indicates that the extracted $I^0_0(Q)$ suggests a greater degree of smearing when the anisotropic resolution function is applied, in contrast to the isotropic resolution function.

%---------------------------------------------
\begin{figure}
\centerline{
  \includegraphics[width =\columnwidth]{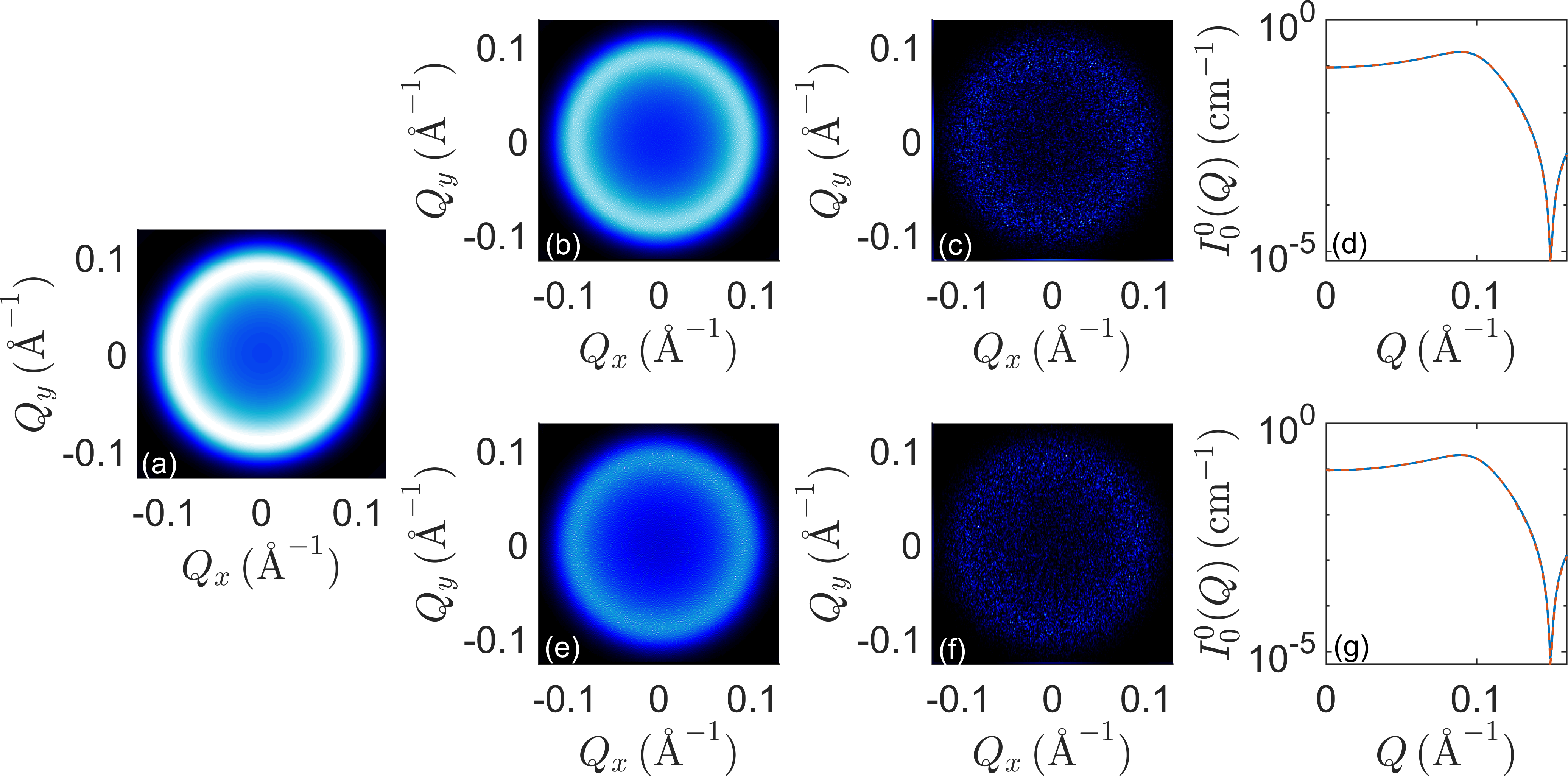}
  %\vspace{0 cm}
}  
\caption{The impact of statistical errors on spectral desmearing is assessed. The desmeared two-dimensional scattering intensities for isotropic and anisotropic resolution functions are shown in Figs.~\ref{fig:5}(b) and~(e). The intensity differences with the ground truth are presented in Figs.~\ref{fig:5}(c) and~(f), demonstrating that discrepancies fall within statistical uncertainties. The residual noise distribution (Fig.~\ref{fig:5}(f)) confirms that the desmearing algorithm corrects for spectral anisotropy. Finally, Fig.~\ref{fig:5}(g) shows that the radially averaged one-dimensional intensity is fully recovered by the desmearing method. For the isotropic resolution, \( dQ_x = dQ_y = 1.0 \times 10^{-3} \,\text{\AA}^{-1} \). For the anisotropic resolution, \( dQ_x = 1.5 \times 10^{-3} \,\text{\AA}^{-1} \) and \( dQ_y = 6.0 \times 10^{-3} \,\text{\AA}^{-1} \).}
\label{fig:5}
\end{figure}
%---------------------------------------------

The impact of statistical errors on spectral desmearing is now clearly assessed. The desmeared two-dimensional scattering intensities, corresponding to isotropic and anisotropic resolution functions, are presented in Figs.~\ref{fig:5}(b) and~(e), respectively. The intensity differences between these desmeared intensities and the ground truth noisy, unsmeared scattering intensity, shown in Fig.~\ref{fig:5}(a), are illustrated in Figs.~\ref{fig:5}(c) and~(f). These results indicate that the discrepancy between the desmeared and smeared intensities falls within the magnitude of the statistical uncertainties. This outcome aligns with expectations, as the numerical accuracy of desmeared intensities cannot exceed the inherent measurement uncertainties. Furthermore, the angular distribution of residual noise in Fig.~\ref{fig:5}(f) demonstrates that the desmearing algorithm effectively corrects for the artificial spectral anisotropy introduced by the anisotropic instrument resolution. Finally, the results in Fig.~\ref{fig:5}(g) confirm that the radially averaged one-dimensional intensity can be fully recovered from the instrument resolution by our proposed desmearing method.

After computationally benchmarking the numerical reliability of our proposed desmearing algorithm using the angularly isotropic scattering intensities of an interacting hard sphere system in its equilibrium state, we now assess its feasibility for desmearing angularly anisotropic 2D scattering intensities, commonly encountered in rheo-SANS experiments of mechanically driven soft materials. To this end, we utilize the Gaussian chain system, which is subject to uniaxial tension and undergoing affine deformation as our testbed. The Gaussian chain is a theoretical model widely employed in polymer physics to describe the behavior of polymer chains in solution~\cite{Yamakawa}. In this model, the polymer chain is represented as a series of connected segments, each corresponding to a monomer or a group of monomers. The chain follows a random walk pattern, where the orientation of each segment is independent of the others. Consequently, the end-to-end distance of the chain is distributed according to a Gaussian distribution, in accordance with the central limit theorem~\cite{Kardar}. 

The Gaussian chain model assumes idealized conditions, such as the absence of excluded volume effects, implying that monomers can overlap without repulsion. Despite these simplifications, the model provides valuable insights into polymer properties, including the contour length, root-mean-square end-to-end distance, and radius of gyration. This makes it a foundational tool for understanding more complex polymer systems.

The coherent neutron scattering intensity, \( I(Q) \), for this system, first derived by Debye~\cite{Debye}, is given by:
\begin{equation}
I(Q) = 2\frac{e^{-x} + x - 1}{x^2},
\label{eq:3.1}
\end{equation}
where \( x \equiv (Q R_g)^2 \) and \( R_g \) is the radius of gyration. 

For the uniaxial affine tension of a Gaussian chain, the corresponding scattering intensity on the \(Q_x\)-\(Q_y\) plane can be expressed as:
\begin{equation}
    I(Q_x, Q_y) = 2 \frac{e^{-y} + y - 1}{y^2}
\end{equation}
where 
\begin{equation}
    y = \left(\frac{Q_x^2}{\gamma^2} + \gamma Q_y^2\right) R_g^2
\end{equation}
and \(\gamma\) is the magnitude of the tensile strain.

\begin{figure}
\centerline{
  \includegraphics[width =\columnwidth]{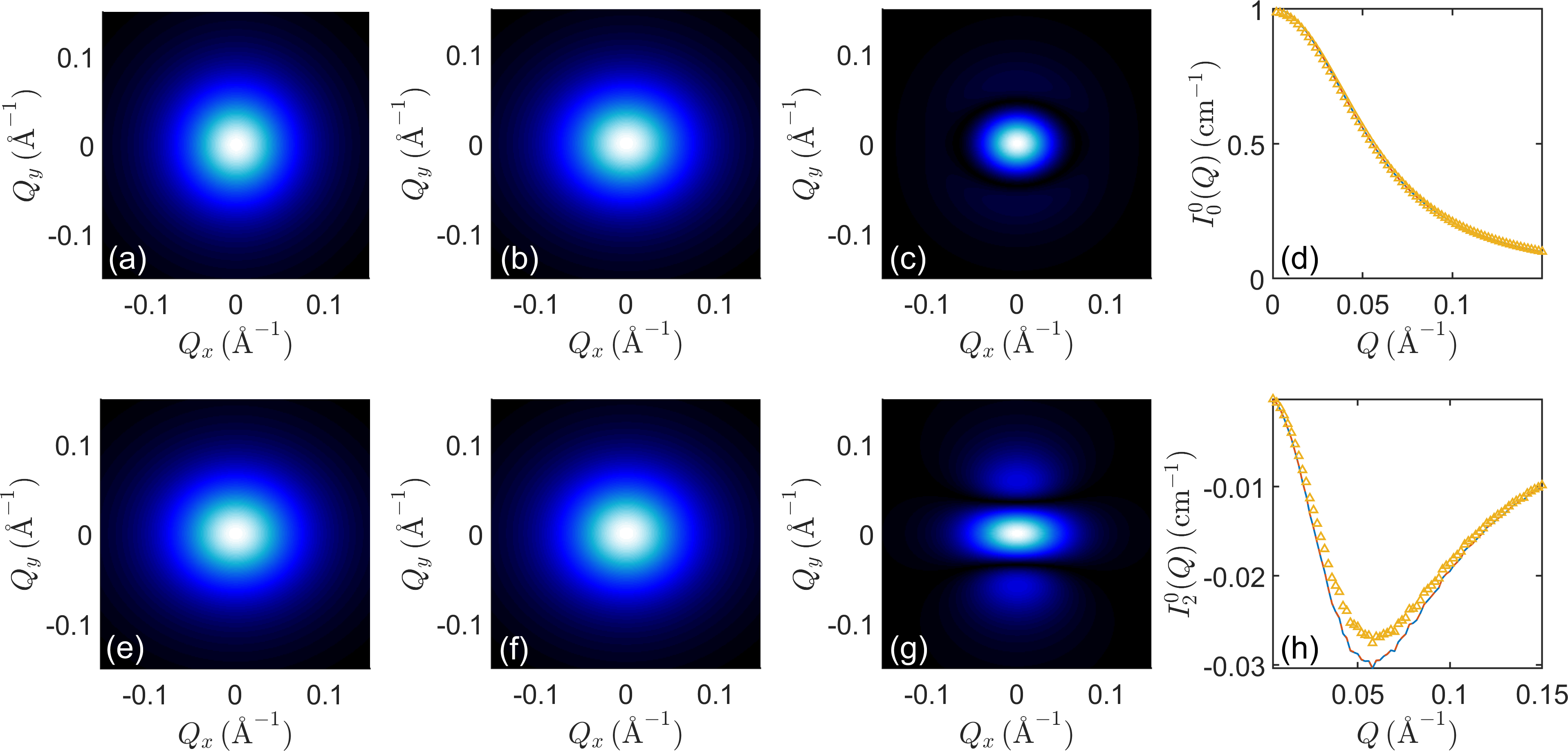}
  %\vspace{0 cm}
}  
\caption{Panel (a) shows the unsmeared \( I(\textbf{Q}) \) of a Gaussian chain in its quiescent state, while Panel (e) shows the scattering intensity of a Gaussian chain under affine stretching with \( \gamma = 0.5 \), displaying stretching along the \( Q_x \)-direction. Panels(b) and~(f) show the anisotropic scattering intensities smeared by isotropic and anisotropic resolution functions, respectively. Although visual inspection reveals no significant differences, the corresponding \( \Delta I \) in Panels(c) and~(g) highlights the smearing effects. Comparison of \( \Delta I \) in Panel (g) with~\ref{fig:6}(c) shows the enhanced spectral anisotropy due to the anisotropic resolution function. Panel (d) shows that while \( I^0_0(Q) \) remains unaffected, \( I^2_0(Q) \) extracted from Panel (f) exhibits significant deviation around the first minimum at \( 0.05 \, \rm{\AA^{-1}} \), as shown in Panel (h). For the isotropic resolution, \( dQ_x = dQ_y = 1.0 \times 10^{-3} \,\text{\AA}^{-1} \). For the anisotropic resolution, \( dQ_x = 1.5 \times 10^{-3} \,\text{\AA}^{-1} \) and \( dQ_y = 6.0 \times 10^{-3} \,\text{\AA}^{-1} \).}
\label{fig:6}
\end{figure}

In a manner similar to the analysis presented in Fig.~\ref{fig:2}, we first examine noise-free scattering intensities, with the results shown in Fig.~\ref{fig:6}. Fig.~\ref{fig:6}(a) displays the unsmeared \( I(\textbf{Q}) \) of a Gaussian chain in its quiescent state, while Fig.~\ref{fig:6}(e) shows the scattering intensity of a Gaussian chain that is affinely stretched along the \( y \)-direction in real space with a strain of \( \gamma = 0.5 \). As a result, the scattering intensity exhibits stretching along the \( Q_x \)-direction.

Figs.~\ref{fig:6}(b) and~\ref{fig:6}(f) present the anisotropic scattering intensities obtained from the unsmeared intensity in Fig.~\ref{fig:6}(e), smeared by isotropic and anisotropic resolution functions, respectively. While visual inspection reveals no substantial differences between these two cases, the corresponding \( \Delta I \), shown in Figs.~\ref{fig:6}(c) and~\ref{fig:6}(g), clearly highlights the smearing effects due to the instrument resolution. A comparison of \( \Delta I \) in Fig.~\ref{fig:6}(g) with that in Fig.~\ref{fig:6}(c) demonstrates the enhanced spectral anisotropy induced by the anisotropic resolution function. Furthermore, as shown in Fig.~\ref{fig:6}(d), where the red curve, yellow symbol, and blue curve represent results extracted from panels~(e), (d), and (f) respectively, the extracted \( I^0_0(Q) \) remains unaffected by the instrument resolution. In contrast, the \( I^2_0(Q) \) extracted from Fig.~\ref{fig:6}(f) exhibits considerable deviation around the first minimum at approximately \( 0.05 \, \rm{\AA^{-1}} \), as indicated by Fig.~\ref{fig:6}(h).

%---------------------------------------------
%---------------------------------------------

The observation underscores the importance of demearing the instrument resolution in rheo-SANS experiments. One of the key pieces of information for mechanically driven polymer systems is the orientational distribution function (ODF), which characterizes the degree of alignment. In spherical harmonic expansion analysis, \( I^2_0(Q) \) plays a crucial role in determining the ODF of stretched polymers \cite{Huang2019, Huang2021, Huang2025}. Without resolution correction, it is evident that the inverted ODF derived from untreated smeared \( I^2_0(Q) \) will lead to erroneous results.

%---------------------------------------------
\begin{figure}
\centerline{
  \includegraphics[width =\columnwidth]{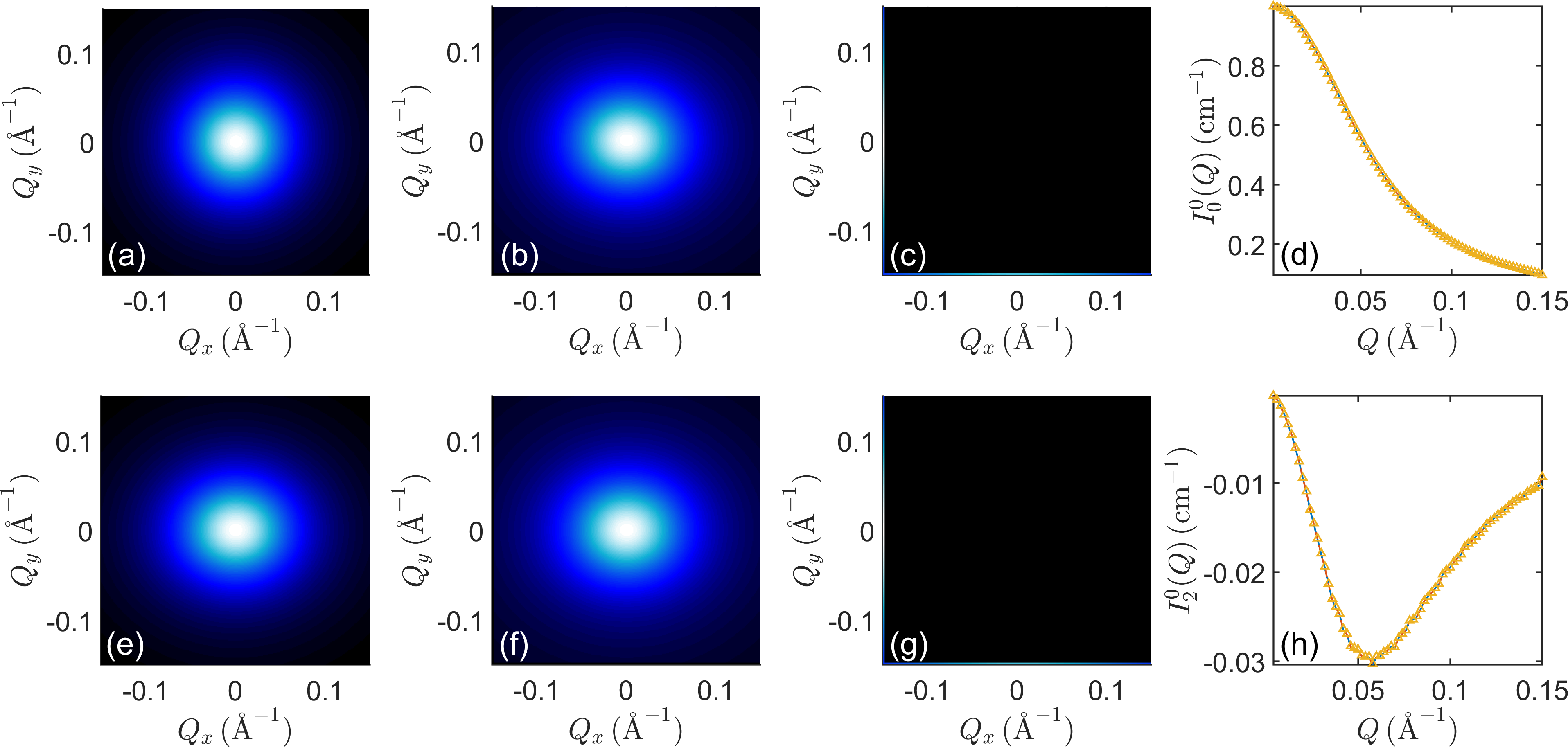}
  %\vspace{0 cm}
}  
\caption{We evaluate the feasibility of our algorithm for desmearing the effects of instrument resolution on noise-free anisotropic scattering intensities. The panel arrangement follows that of Figs.~\ref{fig:3} and~\ref{fig:6}. The results, presented as \( \Delta I \) in Figs.~\ref{fig:7}(c) and~\ref{fig:7}(g), and \( I^0_0(Q) \) in Fig.~\ref{fig:7}(d) and \( I^2_0(Q) \) in Fig.~\ref{fig:7}(h), demonstrate near-perfect spectral restoration. These findings validate the effectiveness of our desmearing algorithm in removing resolution-induced smearing from anisotropic scattering intensities. For the isotropic resolution, \( dQ_x = dQ_y = 1.0 \times 10^{-3} \,\text{\AA}^{-1} \). For the anisotropic resolution, \( dQ_x = 1.5 \times 10^{-3} \,\text{\AA}^{-1} \) and \( dQ_y = 6.0 \times 10^{-3} \,\text{\AA}^{-1} \).
}
\label{fig:7}
\end{figure}
%---------------------------------------------

Next, as depicted in Fig.~\ref{fig:7}, we assess the performance of our algorithm in mitigating the impact of instrument resolution on noise-free anisotropic scattering intensities. The panel layout follows that of Figs.~\ref{fig:3} and~\ref{fig:6}. The results, presented as \( \Delta I \) in Figs.~\ref{fig:7}(c) and~\ref{fig:7}(g), \( I^0_0(Q) \) in Fig.~\ref{fig:7}(d), and \( I^2_0(Q) \) in Fig.~\ref{fig:7}(h), exhibit near-perfect spectral restoration. These findings affirm the efficacy of our desmearing algorithm in eliminating resolution-induced smearing from anisotropic scattering intensities.

\begin{figure}
\centerline{
  \includegraphics[width =\columnwidth]{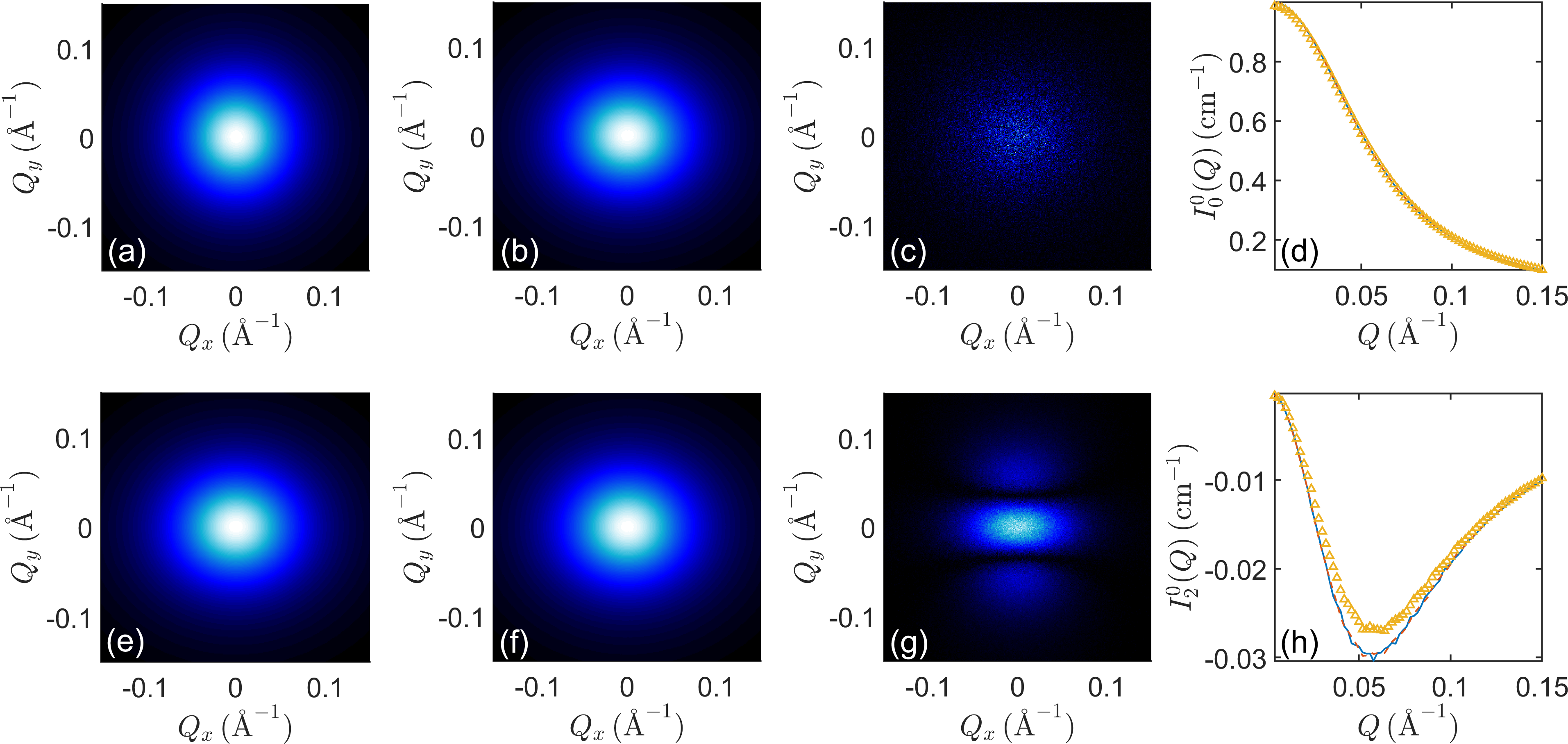}
  %\vspace{0 cm}
}  
\caption{The impact of instrument resolution on the blurring of distinct features in scattering intensities, compounded by statistical uncertainties, is illustrated here. The panel layout follows that of Figs.~\ref{fig:4} and~\ref{fig:6}. A comparison of \( \Delta I \) in Panels (c) and (g) reveals that anisotropic smearing induces significantly greater spectral distortion than isotropic smearing. As shown in Panels (d) and (h), while the corresponding \( I^0_0(Q) \) in Panel (d) remains largely unaffected, a noticeable deviation between \( I^0_0(Q) \) extracted from Panels (f) (yellow symbols) and (e) (red curve) is evident, as illustrated in Panel (h). For the isotropic resolution, \( dQ_x = dQ_y = 1.0 \times 10^{-3} \,\text{\AA}^{-1} \). For the anisotropic resolution, \( dQ_x = 1.5 \times 10^{-3} \,\text{\AA}^{-1} \) and \( dQ_y = 6.0 \times 10^{-3} \,\text{\AA}^{-1} \).}
\label{fig:8}
\end{figure}

Fig.~\ref{fig:8} illustrates the effect of instrument resolution on the blurring of distinctive features in scattering intensities, further exacerbated by statistical uncertainties. The panel layout follows that of Figs.~\ref{fig:4} and~\ref{fig:6}. A comparison of \( \Delta I \) in Figs.~\ref{fig:8}(c) and~\ref{fig:8}(g) reveals that anisotropic smearing causes significantly greater spectral distortion than isotropic smearing. As seen in Figs.~\ref{fig:6}(d) and~\ref{fig:6}(h), while the corresponding \( I^0_0(Q) \) in Fig.~\ref{fig:8}(d) remains largely unaffected, a clear deviation between \( I^0_0(Q) \) extracted from Fig.~\ref{fig:8}(f) (yellow symbols) and that from Fig.~\ref{fig:8}(e) (red curve) is evident, as shown in Fig.~\ref{fig:8}(h).

%---------------------------------------------
%---------------------------------------------

\begin{figure}
\centerline{
  \includegraphics[width =\columnwidth]{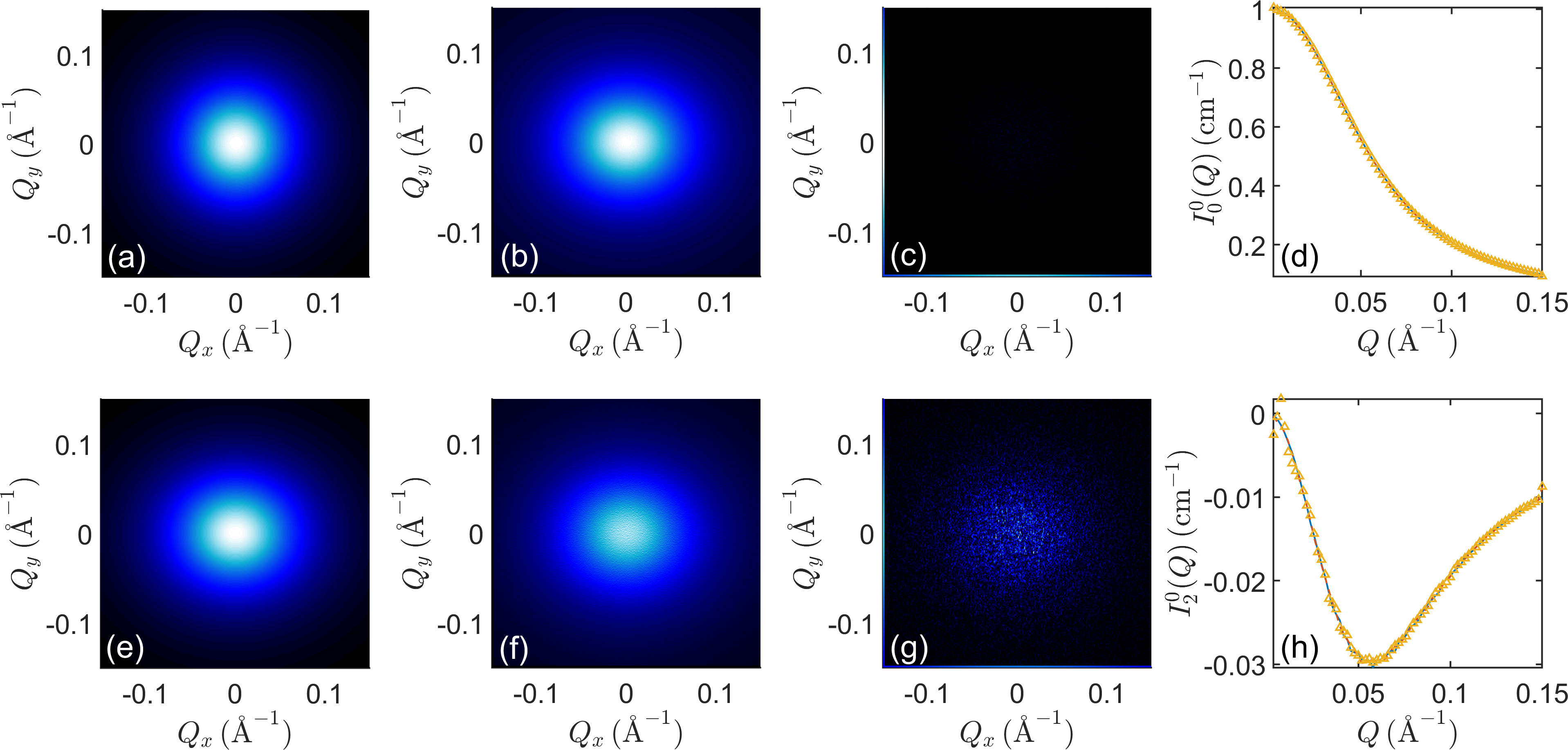}
  %\vspace{0 cm}
}  
\caption{To validate the desmearing algorithm, 0.d1\% Gaussian noise is added to the scattering intensities from Panels~\ref{fig:8}(e), (b), and (f), with the resulting noisy data shown in Panels~\ref{fig:9}(e), (b), and (f). A comparison of the \( \Delta I \) values in Panels~\ref{fig:8}(c) and~\ref{fig:8}(g), as well as the extracted \( I^0_0(Q) \) and \( I^2_0(Q) \) in Panels~\ref{fig:8}(d) and~\ref{fig:8}(h), demonstrates that discrepancies are within the expected uncertainty range, confirming the algorithm's accuracy.
}
\label{fig:9}
\end{figure}

Finally, we evaluate the numerical accuracy of our desmearing algorithm in removing instrument resolution effects from noisy anisotropic scattering spectra. To simulate experimentally obtained spectra, Gaussian noise corresponding to 1\% statistical uncertainty is added to the scattering intensities from Figs.~\ref{fig:8}(e), (b), and (f) at each sampled $\textbf{Q}$ point. The resulting noisy data are presented in Figs.~\ref{fig:9}(e), (b), and (f), respectively. By comparing the corresponding \( \Delta I \) values from Figs.~\ref{fig:8}(c) and~\ref{fig:8}(g), along with the extracted \( I^0_0(Q) \) and \( I^2_0(Q) \) shown in Figs.~\ref{fig:8}(d) and~\ref{fig:8}(h), we confirm that the magnitude of the quantitative discrepancies remains within the range of uncertainties. Thus, the feasibility of the algorithm is successfully validated.

%---------------------------------------------
%---------------------------------------------

\section{Experimental Validation}
\label{sec:Computational Benchmark}

To demonstrate the practicality of the desmearing algorithm in rheo-SANS experiments, we analyzed the SANS intensity, $I(Q)$, of nanocomposites with a composition of 20 wt\% Pluronic F127 and 2 wt\% nanoclay. The preparation protocol was provided in detail elsewhere \cite{Jin1, Jin2}.

First, the nanoclay suspension was prepared by dispersing a given weight of dry nanoclay powder (Laponite, BYK Additives Inc., Gonzales, TX) into $D_2O$ at room temperature and mixing thoroughly to ensure complete hydration. Then, Pluronic F127 powder (P2443, Sigma-Aldrich, Burlington, MA) was added to the suspension at a specific concentration and mixed at 4$^{\circ}$C until fully dissolved. Finally, the prepared nanocomposite was degassed using a centrifuge to remove entrapped air bubbles, preventing any adverse effects on the SANS measurements. The bubble-free nanocomposite was stored in a sealed container and aged at 4$^{\circ}$C for one day before use. 

The microscopic structure of the sheared nanocomposites solution was investigated through flow-SANS measurements, using a shear cell configured in Couette flow geometry. The experimental design focused on three principal directions aligned with the applied shear flow: the flow direction (\(\mathbf{v}\), denoted as the 1-direction), the velocity gradient direction (\(\nabla \mathbf{v}\), denoted as the 2-direction), and the vorticity direction (\(\mathbf{x} = \nabla \times \mathbf{v}\), denoted as the 3-direction). SANS measurements were performed in the flow–vorticity plane (\(\mathbf{v}-\nabla \times \mathbf{v}\), or the 1–3 plane). Throughout the measurements, the temperature was maintained at 25~$^{\circ}$C.

%---------------------------------------------
\begin{figure}
\centerline{
  \includegraphics[width =\columnwidth]{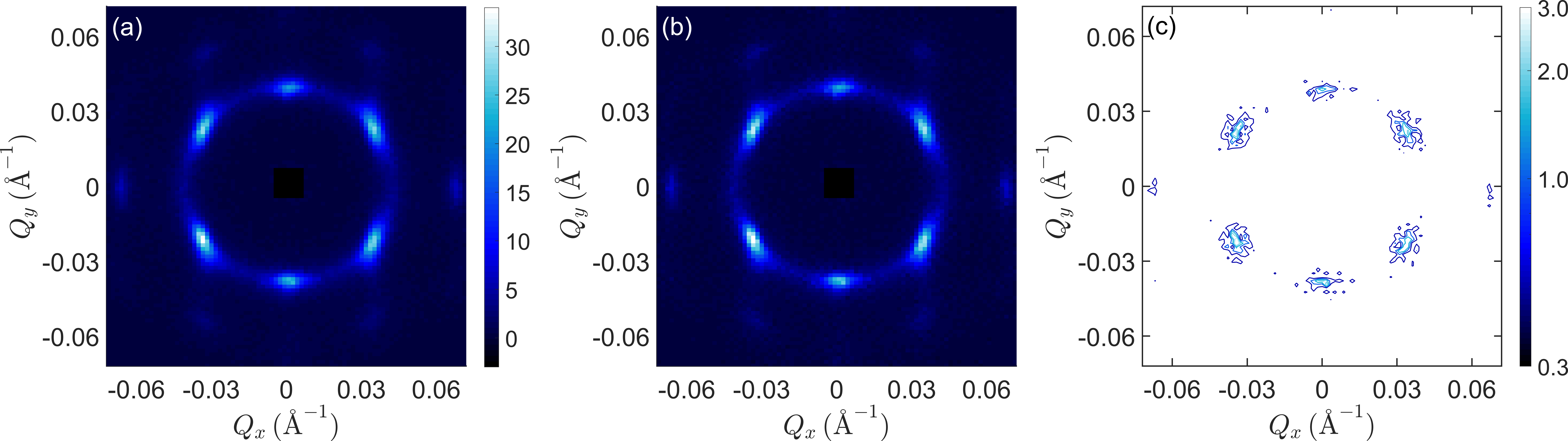}
  %\vspace{0 cm}
}  
\caption{Scattering intensity, \(I(\textbf{Q})\), of the F127-nanoclay nanocomposites system subjected to a shear rate of \(1000~\mathrm{s}^{-1}\). (a) The prominent first correlation peak centered at approximately \(0.03~\mathrm{\AA}^{-1}\) and a second, less pronounced correlation peak at around \(0.06~\mathrm{\AA}^{-1}\) are observed. The sixfold symmetry in the scattering pattern suggests shear-induced crystallization of spherical micelles with face-centered cubic (FCC) symmetry projected onto the 1--3 plane. (b) The desmeared \(I(\textbf{Q})\). (c) The intensity difference between the smeared and desmeared data, showing that key features such as peak heights, widths, and the depth and spread of zeros can be reliably inverted. The intensity difference is approximately \(3~\mathrm{cm}^{-1}\), with a 10\% correction in the intensity of the first correlation peak based on the smeared intensity in panel (a). The resolution function used here is identical to those in panels (a) and (b) of Fig. 1. }

\label{fig:10}
\end{figure}
%---------------------------------------------

The SANS experiments were conducted at the EQSANS beamline of SNS. Data acquisition used a 4 m sample-to-detector distance with a wavelength band defined by a minimum wavelength of 2.5~\(\mathrm{\AA}\). A source aperture of 25 mm diameter (positioned 10.08 m from the source) was used in conjunction with a sample aperture of 10 mm diameter placed directly in front of the sample position (14.30 m from the source). Data corrections accounted for detector noise and pixel-dependent sensitivity, and background scattering from an empty cell was subtracted. During data reduction, time-of-flight events were binned using a constant wavelength step of 0.1~\(\mathrm{\AA}\).

Fig.~\ref{fig:10}(a) shows the scattering intensity, \(I(\textbf{Q})\), of the F127-nanoclay nanocomposites system subjected to a shear rate of \(1000~\mathrm{s}^{-1}\). The first correlation peak, centered at approximately \(0.03~\mathrm{\AA}^{-1}\), is prominent, while a second, less pronounced correlation peak appears around \(0.06~\mathrm{\AA}^{-1}\). The distinct sixfold symmetry observed in the scattering pattern clearly indicates shear-induced crystallization of the constituent spherical micelles, with a face-centered cubic (FCC) symmetry projected onto the 1--3 plane. Fig.~\ref{fig:10}(b) presents the desmeared \(I(\textbf{Q})\). The intensity difference shown in Fig.~\ref{fig:10}(c) illustrates that key features of \(I(\textbf{Q})\)—such as the peak heights and widths, as well as the depth and spread of the zeros—can be reliably inverted. Based on the color changes in Fig.~\ref{fig:10}(c), the intensity difference is approximately \(3~\mathrm{cm}^{-1}\). Given the peak height of \(I(\textbf{Q})\) in the smeared intensity (Fig.~\ref{fig:10}(a)), a 10\% correction in the intensity of the first correlation peak is observed.

Our algorithm also offers a distinct advantage, as it enables straightforward tracing of statistical fluctuations inherent to the measured scattering cross sections. This capability allows us to confidently assign experimental uncertainties to the reconstructed \(I(\textbf{Q})\).

\section{Conclusion}
\label{sec:Conclusion}

In this study, we introduced a novel desmearing methodology based on the central moment expansion technique to address the resolution smearing effects in two-dimensional small-angle neutron scattering (SANS) data. This work builds on prior success in desmearing one-dimensional scattering data and extends the framework to tackle the unique challenges posed by anisotropic scattering spectra commonly observed in rheo-SANS experiments. These challenges include directional variations in resolution smearing and the computational complexity associated with higher-dimensional data.

Through computational benchmarks, we demonstrated the numerical accuracy and reliability of the proposed desmearing algorithm in recovering the true scattering intensities of both isotropic and anisotropic systems. Using test cases such as interacting hard-sphere fluids and Gaussian chains under quiescent and deformed conditions, the method effectively resolved resolution-induced spectral distortions while maintaining fidelity to ground truth data. Importantly, the algorithm performed robustly even in the presence of statistical noise, highlighting its practical relevance for experimental applications.

Experimental validation further underscored the feasibility and effectiveness of the algorithm. By applying the desmearing method to rheo-SANS data of a nanocomposites solution under shear flow, we successfully corrected resolution-induced distortions and recovered key structural features, such as peak heights, widths, and the depth of zeros. The sixfold symmetry observed in the scattering pattern highlighted the algorithm's ability to reconstruct shear-induced crystallization of micelles with face-centered cubic symmetry. Additionally, the desmearing process allowed for the accurate assignment of experimental uncertainties, reinforcing its robustness for real experimental applications.

Our results underscore the critical role of resolution correction in obtaining accurate structural parameters from anisotropic SANS data. Without proper resolution correction, the analysis of orientational distribution functions and other key structural metrics may yield erroneous interpretations, hindering the understanding of material behavior under dynamic conditions. This generalized desmearing methodology eliminates the need for predefined structural models, enhancing flexibility and applicability across a wide range of systems.

Future work will focus on extending the methodology to incorporate additional experimental uncertainties, such as detector noise and non-Gaussian resolution functions, and further validating the algorithm with diverse rheo-SANS datasets. Ultimately, this work represents a significant step forward in the quantitative analysis of anisotropic SANS data, enabling deeper insights into the relationship between material microstructure and macroscopic properties.

\clearpage

     % Appendices appear after the main body of the text. They are prefixed by
     % a single \appendix declaration, and are then structured just like the
     % body text.

%\appendix
%\section{Appendix title}

%\subsection{Title}

%\subsubsection{Title}

     %-------------------------------------------------------------------------
     % The back matter of the paper - acknowledgements and references
     %-------------------------------------------------------------------------

     % Acknowledgements come after the appendices

\ack{Acknowledgements}

This research at ORNL's Spallation Neutron Source was sponsored by the Scientific User Facilities Division, Office of Basic Energy Sciences, U.S. Department of Energy. This research was sponsored by the Laboratory Directed Research and Development Program of Oak Ridge National Laboratory, managed by UT-Battelle, LLC, for the U. S. Department of Energy. The beam time was allocated to EQSANS on proposal number IPTS-32804.1. G.R.H. was supported by the National Science and Technology Council in Taiwan with Grant No. NSTC 111-2112-M-110-021-MY3. Y.S. was supported by the U.S. Department of Energy (DOE), Office of Science, Office of Basic Energy Sciences, Materials Sciences and Engineering Division, under Contract No. DE-AC05-00OR22725.

\clearpage

\bibliographystyle{iucr}
\bibliography{references}

@article{Hayter1984,
  title={Use of Viscous Shear Alignment To Study Anisotropic Micellar Structure by Small-Angle Neutron Scattering},
  author={Hayter, J. B. and Penfold, J.},
  journal={J. Phys. Chem.},
  volume={88},
  pages={4589-4593},
  year={1984}
}

@article{Hayter1987,
  title={A Small-angle Neutron Scattering Investigation of Rod-like Micelles aligned by Shear Flow},
  author={Cummins, P. G. and Staples, E. and Hayter, J. B. and Penfold, J.},
  journal={J. Chem. Soc., Faraday Trans. 1},
  volume={83},
  pages={2773–2786},
  year={1987}
}

@article{Huang2023,
  title={Desmearing Small-Angle Scattering Data by Central Moment Expansions},
  author={Huang, G.-R. and Tung, C.-H. and Chen, M.-Z. and Porcar, L. and Shinohara, Y. and Wildgruber, C. U. and Do, C. and Chen, W.-R.},
  journal={J. Appl. Cryst.},
  volume={56},
  pages={1537–1543},
  year={2023}
}

@article{OZ,
  author = {L. S. Ornstein and F. Zernike}, 
  title ={Accidental deviations of density and opalescence at the critical point of a single substance},
  journal = {Proc. Akad. Sci.},
  volume = {17}, 
  pages = {793-806},
  year = {1914},
}

@article{PY,
  title={Analysis of Classical Statistical Mechanics by Means of Collective Coordinates},
  author={Percus, Jerome K. and Yevick, George J.},
  journal={Phys. Rev.},
  volume={110},
  number={1},
  pages={1--13},
  year={1958},
  publisher={APS}
}

@article{Baxter1,
  title={ORNSTEIN-ZERNIKE RELATION FOR A DISORDERED FLUID},
  author={Baxter, Rodney J.},
  journal={Aust. J. Phys.},
  volume={21},
  pages={563--569},
  year={1968},
  publisher={Taylor \& Francis}
}

@article{Baxter2,
  title={Ornstein-Zernike Relation and Percus-Yevick Approximation for Fluid Mixtures},
  author={Baxter, Rodney J.},
  journal={J. Chem. Phys.},
  volume={52},
  number={9},
  pages={4559--4562},
  year={1970},
  publisher={American Institute of Physics}
}

@BOOK{Hansen,
  title = {Theory of Simple Liquids with Applications to Soft Matter, 4th Ed.},
  publisher = {Academic Press},
  year = {2013},
  author = {J.-P. Hansen and I. R. McDonald},
  address = {Amsterdam}
}

@article{Wertheimt,
  title={EXACT SOLUTION OF THE PERCUS- YEVI CK INTEGRAL EQUATION FOR HARD SPHERES},
  author={Wertheimt, M. S.},
  journal={Phys. Rev. Lett.},
  volume={10},
  number={8},
  pages={321--323},
  year={1963},
  publisher={APS}
}

@article{Debye,
  title={Molecular-weight determination by light scattering},
  author={Debye, P.},
  journal={J. Phys. Colloid Chem.},
  volume={18},
  pages={18--32},
  year={1947}
}

@book{Kardar,
  author = {Kardar, M.},
  title  =  {Statistical Physics of Particles},
  publisher = {Cambridge University Press},
  address={Cambridge},
  year   = {2007}       
}

@book{Yamakawa,
  author = {Yamakawa, H.},
  title  =  {Modern theory of polymer solutions},
  publisher = {Harper \& Row},
  address={Manhattan},
  year   = {1971}       
}

@book{Born_Wolf,
    author = {Born, M. and Wolf, E.},
    title = {Principles of Optics},
    publisher = {Cambridge University Press},
    edition = {7th edition},
    address = {UK},
    year = {1999}
}

@ARTICLE{Huang2019,
  author  = {Huang, G.-R. and Wang, Y. and Do, C. and Shinohara, Y. and Egami, T. and Porcar, L. and Liu, Y. and Chen, W.-R.},
  title   = {Orientational Distribution Function of Aligned Elongated Molecules and Particulates Determined from Their Scattering Signature},
  journal = {ACS Macro. Lett.}, 
  volume  = {8},
  year    = {2019},
  pages   = {1257--1262}
}

@ARTICLE{Huang2021,
  author  = {Huang, G.-R. and Carrillo, J.-M. and Wang, Y. and Do, C. and Porcar, L. and Sumpter, B. G. and Chen, W.-R.},
  title   = {An exact inversion method for extracting orientation ordering by small-angle scattering},
  journal = {Phys. Chem. Chem. Phys.}, 
  volume  = {23},
  year    = {2021},
  pages   = {4120--4132}
}

@article{Huang2025,
  author = {Guan-Rong Huang and Lionel Porcar and Ryan P. Murphy and Yuya Shinohara and Yangyang Wang and Jan-Michael Carrillo and Bobby G. Sumpter and Chi-Huan Tung and Changwoo Do and Wei-Ren Chen},
  title = {Elongated particles in flow: Commentary on small angle scattering investigations},
  journal = {arXiv preprint arXiv:2501.03976},
  year = {2025}
}

@article{Jin1,
  author = {W. Hua and K. Mitchell and L. S. Kariyawasam and C. Do and J. Chen and L. Raymond and N. Valentin and R. Coulter and Y. Yang and Y. Jin},
  title = {Three-dimensional printing in stimuli-responsive yield-stress fluid with an interactive dual microstructure},
  journal = {ACS Appl. Mater. Interfaces}, 
  volume  = {14},
  year    = {2022},
  pages   = {39420--39431}
}

@article{Jin2,
  author = {C. Zhang and W. Hua and K. Mitchell and L. Raymond and F. Delzendehrooy and L. Wen and C. Do and J. Chen and Y. Yang and G. Linke and Z. Zhang and M. Krishnan and M. Kuss and R. Coulter and E. Bandala and Y. Liao and B. Duan and D. Zhao and G. Chai and Y. Jin},
  title = {Multiscale embedded printing of engineered human tissue and organ equivalents},
  journal = {Proc. Natl. Acad. Sci. U.S.A.}, 
  volume  = {12},
  year    = {2024},
  pages   = {e2313464121}
}

     %-------------------------------------------------------------------------
     % TABLES AND FIGURES SHOULD BE INSERTED AFTER THE MAIN BODY OF THE TEXT
     %-------------------------------------------------------------------------

\end{document}